# Connecting the Dots of Knowledge in Agile Software Development

Raquel Ouriques, *Blekinge Institute of Technology, Sweden*

Tony Gorschek, *Blekinge Institute of Technology, Sweden, and fortiss, Germany*

Daniel Mendez, *Blekinge Institute of Technology, Sweden, and fortiss, Germany*

Fabian Fagerholm, *Aalto University, Finland*

*Abstract*—This article discusses the importance of managing knowledge as a resource due to its great potential to create economic value. We detail the types of knowledge resources, the challenges associated with their management, and potential solutions to maximise their utility. Our contribution is based on empirical studies performed in an industry context.

Knowledge is the fundamental resource for executing software development activities [1], put under the spotlight when agile principles started to be incorporated by companies. The shift from traditional methods, which included extensive planning and heavy documentation, now emphasises flexibility and focus on people and their knowledge to deal with requirements, design, and implementation evolving iteratively [2].

Collaboration and communication became the critical facilitator for people engaging in creative gatherings for knowledge sharing within self-organised teams [3]. However, knowledge, as an intangible resource, is difficult to reproduce and manage, requiring significant effort to understand in terms of what should remain tacit, and what should be explicitly captured in artefacts, like documentation.

The importance of knowledge as a resource rests on its great potential to create economic value when combined with other people's knowledge, which in turn provides the means for companies to capitalise on it. In this article, we'll talk about why it is so challenging to manage knowledge, understand the proportions of knowledge that remain intangible, and provide insights into how we can effectively manage knowledge stored in artefacts. This is complicated by these key factors:

1) Knowledge belongs to people.
2) Knowing what knowledge should be added to artefacts is complex.

3) Managing knowledge in artefacts requires more coordination than is obvious.

We will also discuss types of knowledge, types of artefacts, and what to add to artefacts. This work is based on studies carried out in industry and reported in detail in peer-reviewed journals [4], [5], [6].

## Knowledge Resources

Before diving into knowledge waters, we first reflect on companies' competitive resources and how they relate to knowledge. Optimisation and efficient allocation of resources are concepts that historically relate to the manufacturing industry. Resource management primarily focuses on planning, allocating people, money, and technology to maximise organisational value [7].

This way of thinking is still valid in the development of software-intensive products and services. However, the proportions of the type of resources have changed. In the software industry, co-workers are the critical and most important resource, as their knowledge can generate sustained competitive advantages for companies.

An intangible portion of co-workers' knowledge, which we refer to as **Knowledge-based Resources** (KBRs) (see illustration in Figure 1), is revealed when stored and applied through, e.g., prototypes, architectural descriptions, requirements, and other artefacts. As this knowledge is typically stored in artefacts (documents) or transformed into products, they become **Property-based Resources** (PBRs) as companies own them and can generate profit utilising them [8].

However, a large proportion of these KBRs remain intangible, and are, thus, not realised into PBRs. There might be several reasons for this. The domain of the company might be moving too fast to create PBRs and a PBRs might become outdated too fast to warrant the formalisation investment. Another possible scenario is the complexity of generating PBRs out of KBRs that are usable and useful for others. Knowledge in terms of KBRs are thus a reality that companies live with, and they strongly support software development and enable companies.

## From Artefacts to People

Incorporating agile principles into software engineering shifted the role of artefacts. Informal communication became focus, with the promise of more flexibility when shuffling priorities and re-arranging workload.

Knowledge-based resources were considered essential in this environment as they can contribute significantly to innovation and companies' performance. *KBRs often manifest as specific skills, including technical, creative, integrative and collaborative skills* [8], which we describe and map to the drives, strategies and mechanisms to deal with changes in agile contexts in Figure 2.

The importance of these resources and focus vary depending on the industry domain [9]. Even software companies will have different levels of priority. However, something they all have in common is constant changes. So, what role do KBRs play when companies need to adapt to change in agile contexts?

First, we must understand where these **changes** come from. They are **driven** by market changes or internal transformation (see A in Figure 1). Market changes refer to all external events that force companies to adapt, i.e., technological breakthroughs, new business opportunities, and customer demands. In these scenarios, co-workers utilise their skills and technical capabilities to, for example, analyse the impact of changes on current products, also foreseeing how the product should evolve to match future technologies.

Changes can also spark internally, i.e., re-organisation for efficiency, agile transformation, and product innovation processes. These changes will trigger KBRs, which play a strategic role in incorporating and adapting to changes. Herewith, we can say that KBRs:

1. Help examine internal and external variables affecting the company and define how to respond and incorporate potential changes.

Second, as the changes start to be assimilated, companies must create **strategies** (see B in Figure 1) to disseminate new arrangements originating from changes. They drive co-workers to achieve particular goals through social collaboration, combining their knowledge to deal with the complexity of the products and coordinate actions. For example, by having the technical skill of product awareness (commonly known as big picture), it is possible to evaluate how a new requirement affects the product and its ramification through product development. As it stands, KBRs support the implementation of strategies to deal with changes by:

2. Enhancing social relations within agile teams, encouraging knowledge sharing to solve complex problems associated with software development.

3. Perceiving the changes and their ramifications through product development and adequately adjusting tasks and resources.

Software companies also utilise different **mechanisms** (see C in Figure 1) to ensure that changes are fully implemented. Effective agile teams is one such mechanism, but to achieve change, companies must understand the key aspects of creating a favourable environment where people feel comfortable collaborating. When this mechanism is established, the ability to store and transmit part of the knowledge shared intuitively becomes crucial for companies to *use KBRs to produce PBRs*. By doing this, companies can hold ownership of the co-workers' knowledge as it is stored in artefacts protected by property rights. KBRs support these mechanisms to:

4. Understand what characteristics are relevant in setting up effective agile teams.

5. Support identifying what knowledge needs to be transformed into PBRs.

It is important to note that even though intuitive actions towards the utilisation of KBRs can lead to successful implementations of changes, a lack of insight into managing knowledge produces **inefficiencies**. The lack of a structured process for capturing and storing relevant knowledge results in an overload of information in the artefacts that constitute the PBRs. In these circumstances, co-workers can carry out meaningless searches, confusion about the location of accurate content, and waste time parsing large artifacts. At last, they can get frustrated because of recurring problems that could have been solved if the information had been concise, timely and correct.

The inefficiencies challenge capturing knowledge and transforming it into PBRs (see D in Figure 1). Although the main focus of software companies adopting agile principles is on people rather than artefacts, the proportion of these artefacts' utilised remains critical for the efficiency of the software development life cycle and the principle and practices associated.

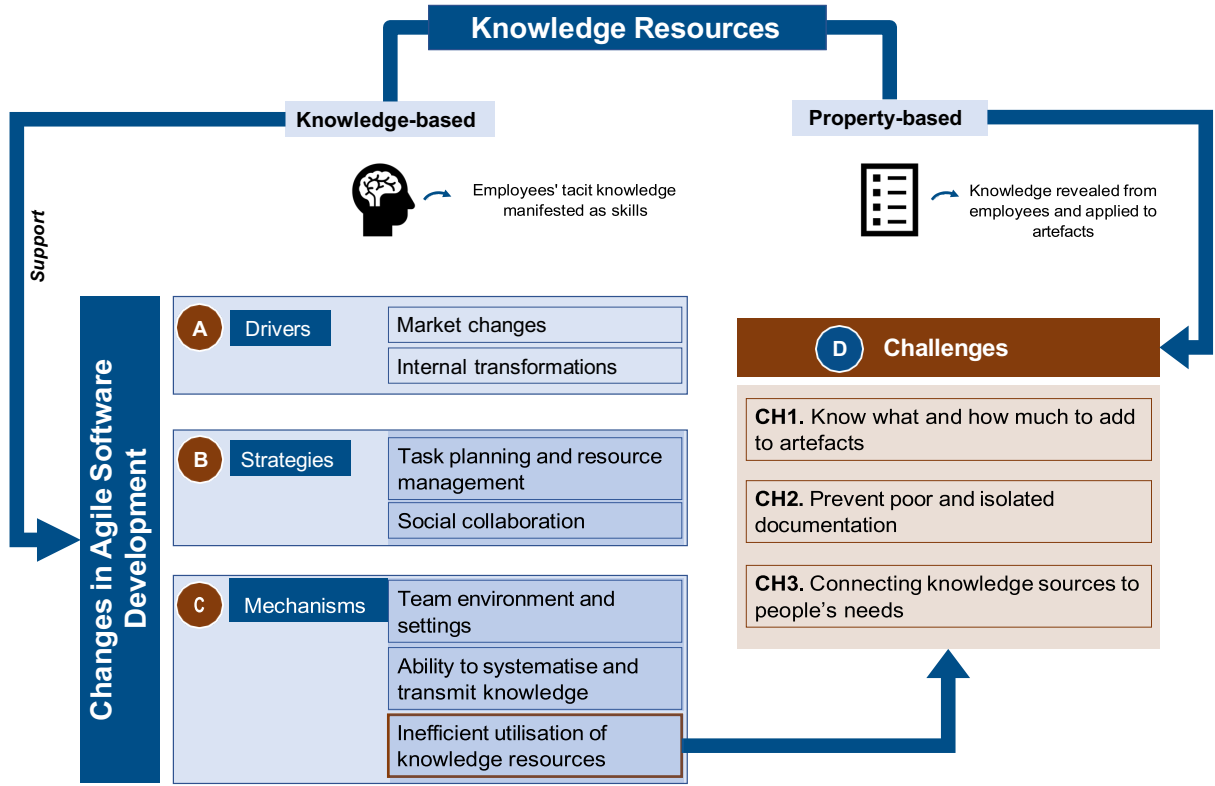

**FIGURE 1.** The types of knowledge resources. The figure shows how knowledge-based resources support changes in agile and the challenges involved in their utilisation to create property-based resources [5].

## Boundaries are Critical

PBRs can be powerful. A single artefact can carry relevant knowledge through different parts (boundaries) of a company. Doing so, they can provide different meanings. For example, a use case that was initially created during requirements engineering can be utilised in planning, design, estimation, and testing activities. In each of those boundaries, people see the use case from their perspective and execute tasks that ultimately contribute to the same larger goal.

The contribution of such artefacts is undoubtedly large. They are flexible in supplying local needs (as per the use case example in Figure 3) while keeping themselves highly structured. Co-workers look at the same information, but what it means to them can differ and usually changes over time.

This type of artefact provides a critical contribu-tion to agile software development contexts. It is a resource when informal communication is not possible. In geographically distributed teams, boundary artefacts are even more important as they can overcome communication barriers and different time zones. Most recently, the pandemic has forced a shift from co-located communication to online formats, bringing many challenges to software teams, including the increased dependence on artefacts.

In some cases, for several reasons, boundary artefacts may not be available or provide accurate content. Co-workers' trust can be affected when such inconsistencies occur, and they become skeptical of using the artefacts [6]. They can avoid using an artefact by searching for information elsewhere, which does not guarantee accuracy either. They can create workarounds to supply that knowledge need, i.e., creating their own versions. The latter becomes dangerous as duplicates can grow, producing uncertainty and even more distrust. Consequences to the software engineering effort can be diverse in such cases and can result in the execution of wrong tasks, utilisation of incorrect information, and setbacks in the development

| SUPPORT | KBRs MANIFESTED AS SKILLS | | | |
|---|---|---|---|---|
| | **TECHNICAL** | **CREATIVE** | **COLLABORATIVE** | **INTEGRATIVE** |
| **DRIVERS** — **Market changes and Internal transformations** | - Ability to know how the current technology should evolve<br>- Ability to balance between business and technical skills<br>- Ability to evaluate the business value in the short term versus long term | - Ability to absorb changes originated from the market<br>- Ability to combine technical capability with market vision<br>- Ability to understand customer value | | - Ability to be ready to absorb changes |
| **STRATEGIES** — **Task planning and resource management** | - Ability to have the perspective of the product to reduce waste<br>- Ability to have product awareness while focusing on small tasks | - Ability to comprehend the implications of change<br>- Ability to understand how to distribute human resources appropriately | | - Ability to accumulate company's experience |
| **Social collaboration** | | - Ability to define formalities to not hinder knowledge creation activities | - Ability to provide socialization processes<br>- Ability to achieve goals by enhancing social collaboration | - Ability to coordinate socialization processes combining employees' personally characteristics<br>- Ability to joint efforts for coordinating the transfer of technical knowledge |
| **MECHANISMS** — **Team environment and settings** | - Ability to understand cognitive processes<br>- Ability to apply suitable practices for social collaboration<br>- Ability to understand knowledge nature to direct the learning strategies. | | - Ability to coexist and interact with different personalities<br>- Ability to constant learn<br>- Ability to know what others know | - Ability to combine interpersonal skills |
| **Ability to systematise and transmit knowledge** | - Ability to identify what knowledge to keep in a systematic way<br>- Ability to promote experience transfer<br>- Ability to represent knowledge efficiently in an artefact<br>- Ability to disseminate architectural knowledge | - Ability to recognize relevant knowledge<br>- Ability to spread awareness of existing knowledge<br>- Ability to perceive artefacts as living documents | - Ability to perceive what produced knowledge could assist other employees | - Ability to balance time allocation to systematize knowledge and time pressure for delivery<br>- Ability to match knowledge needs to the available knowledge<br>- Ability to integrate tools and coordination skills |

**FIGURE 2.** Knowledge-based resources manifested in skills mapped to the drivers, strategies and mechanisms to adapt to changes in agile software development [5].

process (see implications in Figure 3).

Adjusting the boundary artefact to meet co-workers' needs also has challenges (see challenges described in Figure 3) that involve keeping the accuracy of the content and the boundary artefact aligned to its original characteristics. One crucial aspect that has been reported as causing misunderstandings is terminology [10]. When co-workers look at the artefact and do not recognise the terminology, there is a risk of misinterpretation and, consequently, misuse. Even worse is if there is an assumption of a common understanding. In such cases, the benefits of boundary artefacts is undermined. Accuracy and control are also identified challenges that require significant effort to avoid the inefficiency of boundary artefacts.

## Strengthening Property-based Resources

The key to optimising PBRs is to establish formal and structured practices, however it is known that artefacts receive less attention in agile environments [3]. However, as PBRs play a critical role in companies, addressing the challenges involving utilising KBRs to produce PBRs and boundary artefacts is necessary. Here we provide potential solutions targeting the challenges (see Figures 1 and 3) revealed in this article.

This contribution originates from two main research results reported in two papers. One case study investigated the causes and effects of trust in boundary artefacts. The second is a grounded theory study to identify KBRs and explain how they supported changes in agile contexts.

## Challenges for producing PBRs

**CH1. Know what and how much to add to artefacts.** When a structured process for adding content to artefacts is missing, co-workers will fill them according to their preference of what they consider good enough regarding content. The content overload will induce co-workers to meaningless searches where they won't find what they need and lack precision, causing confusion. Conversely, searches won't provide enough knowledge required from co-workers. They will start looking for complementary artefacts to fulfil their needs.

- Solution - CH1. The transformation of the co-workers' knowledge should satisfy their knowledge needs [11]. Identifying co-workers' knowledge needs can be time-consuming, though. However, by being precise and structured, the correct amount of knowledge stored as PBRs tends to be more efficient.

**CH2. Prevent poor and isolated documentation**. One of the consequences of a lack of formal practices is that co-workers can create artefacts within teams without knowing who would benefit from them. Usually, this process is sub-optimal as co-workers are targeting specific internal issues and probably won't maintain that content. In this circumstance, this this practice can overload tools such as Wikis with too much and ill-suited content, which will make it harder, and sometimes confusing, for others to find relevant content.

- Solution - CH2. The creation of official artefacts should be guided by structured procedures that provide at least scope, targeted audience and overall structure [6]. The way the content is structured affects usability and applicability. To help address these concerns, co-workers could think about why people need certain knowledge and how it could be applied. It is worth mentioning that creating such structures does not mean blocking co-workers from creating their artefacts within their teams, bringing rigidity, which goes against agile flexibility. We rather mean that the production of a PBR, which is an investment, should follow an official artefact decision to spread and maintain it.

**CH3. Connecting knowledge sources to people's needs.** Not all PBRs exist in electronic versions such as PDFs or Wiki-based tools. They could, for example, exist in the form of maps, pictures, or sticky notes. It is important to know the format of the planned PBR, and to develop the best way to "connect" it to co-workers who need it.

- Solution - CH3. In software development contexts, most PBRs remain in electronic format, highlighting the relevance of Information and Communication Technologies (ICTs) for connecting co-workers to relevant sources of knowledge [5], [12]. Software companies largely utilise them for coordination, communication, and storing PBRs. However, having ICTs does not guarantee the applicability of the artefacts stored. One should note that, for PBRs, it only means that many co-workers can access them in several parts of the organisation and different locations. The effectiveness, though, relies on the other aspects mentioned above, such as identifying knowledge needs and making sure that each artefact has a deliberate structure and content.

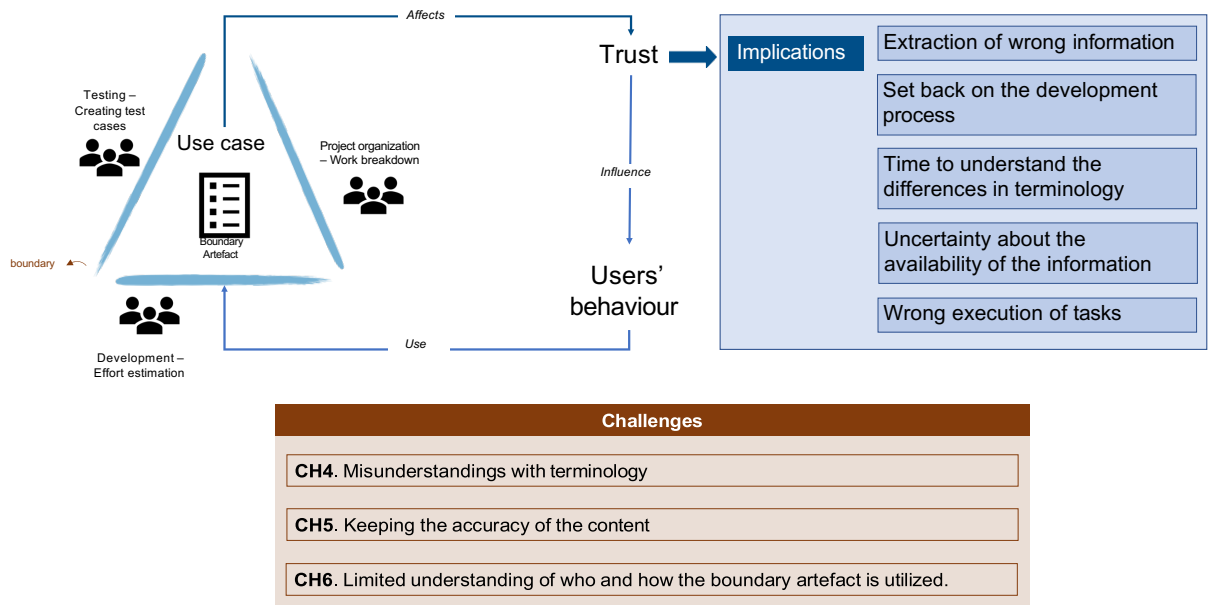

**FIGURE 3.** The cycle of trust in boundary artefacts and how it influences users' behaviour. The figure displays the negative implications of differences in trust and the challenges to managing these artefacts [6].

As we narrow down to specifically boundary artefacts the challenges pertain to the creation and maintenance of boundary artefacts:

**CH4. Misunderstandings with terminology**. Terminology can confuse co-workers if not properly addressed when creating boundary artefacts. Over each boundary, the content can have different meanings and interpretations; if the terms are poorly defined, co-workers can misuse the content.

- Solution CH4. Knowledge can have different levels of novelty, influencing how terminology will be displayed. Carlile [13] introduces three approaches to dealing with terminology. The first refers to generating a common lexicon or taxonomies to address the differences when having a syntactic boundary artefact. The second suggests using cross-functional interactions and boundary translators when addressing semantic boundaries. Finally, pragmatic boundaries focus on combining different knowledge and for that, utilising artefacts that can be jointly created can help, i.e., prototyping.

**CH5. Keeping the accuracy of the content.** When the content is unreliable, co-workers will change their behaviour towards a boundary artefact, creating workarounds, not using them, or making personal "versions". This can happen for many reasons, including a lack of control over the content, ownership issues, and a disconnect between users and creators.

- Solution CH5. To deal with the flexibility of agile environments, a cooperative approach is preferred. Ownership of the boundary artefact can be distributed among users/creators, reducing the workload [14], [12]. For example, in technical artefacts, there could be an owner for the technical part, and another for managing the content. Depending on the size of the artefact and diversity regarding boundaries, ownership can be distributed even more. We also suggest implementing periodic evaluation and feedback processes to include improvements to the process and artifacts as suggested by users, maybe via the retrospective mechanism.

**CH6. Limited understanding of who and how the boundary artefact is utilised**. A key aspect of boundary artefacts is that the content provided in the artifacts reaches predefined users over boundaries. The matching is likely to happen when there have been negotiations and agreements between the different areas of the organization.

- Solution CH6. The simplest solution is to map users' needs, make creators aware, and conduct negotiations [12]. However, this requires maturity in companies when handling boundary artefacts, which is not a guarantee. Most boundary PBRs are well-established artefacts that have been utilized for years despite being inefficient. In this context, we suggest that a diagnosis is made to check if users' content needs are satisfied and map the boundaries for checking terminology differences. As this process requires an investment, it can be more efficient to focus on the severity of negative implications and make the changes gradually focusing on critical PBRs first.

## Reconciling with Traditional Planning

Managing knowledge through PBRs requires more planning and structure than normally associated with agile software development contexts. However, we need to think about the changes software companies have endured since the wide adoption of agile principles, as well as all the principles and practices introduced since then. First, companies have grown and spread development activities across countries. Artefacts have become a key to enabling communication and coordination, as well as being a critical resource for executing tasks.

Second, we have seen lots of evidence showing how the pandemic has changed the ways of working, such as hybrid or completely remote organizations [15]. These changes challenge face-to-face communication and knowledge sharing. As people search for relevant content stored in artefacts, the need to manage knowledge accuracy increases [6].

Traditionally, there has been a dichotomy between agile adoption (albeit not formally expressed in agile principles) and traditional plan-driven organizations. Combining the best of both views seems reasonable. Dybå and Dingsøyr [2] suggest to exploit traditional planning principles in some cases, especially for larger efforts.

Similarly, including more formal planning in managing PBRs does not mean moving back to a documentation-centred approach, but rather managing the creation and maintenance of reliable, predictable and functional artefacts. The main premise of both lean and agile is to "remove waste" and create value. However, overhead, such as coordination and communication through PBRs, is a necessity to be able to create value. Thus, knowledge artefacts that support value creation should be deliberately handled as part of the development process to maximise their utility.

## ACKNOWLEDGMENTS
We would like to acknowledge that this work was supported by the KKS foundation through the S.E.R.T. Research Profile at Blekinge Institute of Technology.

**Raquel Ouriques** is a PhD student in software engineering at the Blekinge Institute of Technology. Her research interest includes knowledge management, boundary artefacts, and trust in inanimate software artefacts.

**Tony Gorschek** is a professor of software engineering at Blekinge Institute of Technology, Karlskrona, Sweden. His research interests include value-based requirements engineering, technology, and product management, process assessment and improvement, real agile and lean, quality assurance, and practical innovation. Further information is available at http://www.gorschek.com.

**Daniel Mendez** Daniel Mendez is professor of software engineering at Blekinge Institute of Technology, Karlskrona, Sweden, and Lead Researcher heading the research division Requirements Engineering at fortiss. His research interests include requirements engineering and software quality. He is further editorial board member for EMSE and JSS where he co-chairs the special tracks Reproducibility & Open Science (EMSE) and In Practice (JSS) respectively. Finally, he is a member of the ACM, the German association of university professors and lecturers, the German Informatics Society, and ISERN. Further information is available at http://www.mendezfe.org.

**Fabian Fagerholm** is assistant professor at the Department of Computer Science, Aalto University, Espoo, Finland. His research interests include organisational and human factors of software development, developer experience, agile software development, and experimental approaches to software product development.